\documentclass[a4paper,11pt]{article}
\usepackage{pos}

\title{Measurement of heavy-flavor production in the high-mass dimuon spectrum in pp collisions at $\sqrt{s}$ = 13 TeV with ALICE}
 
\ShortTitle{Dimuon from heavy-flavor with ALICE}

\author*[a]{Michele Pennisi, on behalf of ALICE Collaboration}

\affiliation[a]{Dipartimento di Fisica, Università degli Studi di Torino, and Sezione INFN Torino\\
Via Pietro Giuria 1, 10125, Torino, Italy}

\emailAdd{michele.pennisi@cern.ch}

\abstract{
Charm and beauty quark production measurement represents a fundamental means to access the initial stage of hadronic collisions. Being produced almost exclusively in initial hard partonic scatterings due to their large masses, $m_{\text{c}}=$ 1.3 $ \text{GeV}/c^2$ and $m_{\text{b}} =$ 4.1 $ \text{GeV}/c^2$, charm and beauty quarks are ideal tools to investigate various QCD aspects.  
In addition, the corresponding measurements in pp collisions represent a baseline for cold nuclear matter studies in p--A collisions as well as for the characterization of the hot and dense interacting nuclear matter, the quark--gluon plasma (QGP), formed in A--A interactions. 

An analysis technique that was little investigated until now at LHC energies to measure $\text{c}\overline{\text{c}}$ and $\text{b}\overline{\text{b}}$ cross sections consists in exploring the high-mass regions of the dilepton invariant mass spectrum, more precisely the two continuum regions between charmonium and bottomonium resonances  $\text{(4} < m_{\mu^{+}\mu^{-}} <\text{ 9 GeV}/c^2$) and above the bottomonium states ($m_{\mu^{+}\mu^{-}}> $ 11 $ \text{GeV}/c^2$). Indeed, these regions are significantly populated by the semileptonic decays of hadron pairs containing charm and beauty quarks.

In these proceedings, the status of this study is shown, discussing the details of the analysis procedure foreseen to measure the charm and beauty cross sections. In particular, the results of PYTHIA8 simulations for pp collisions at $\sqrt{s}$ = 13 TeV in the rapidity region 2.5 < \textit{y} < 4 covered by the ALICE muon spectrometer, crucial to address the different components to the dimuon yield, are presented.

}

\FullConference{%
  The Tenth Annual Conference on Large Hadron Collider Physics - LHCP2022\\
  16-20 May 2022\\
  online
}

\begin{document}

\maketitle
\section{Introduction}
In the context of heavy-flavor (HF) analyses,charmonia ($\text{c}\overline{\text{c}}$) or bottomonia ($\text{b}\overline{\text{b}}$), i.e. bound states of charm quarks and antiquarks, or beauty quarks and antiquarks, respectively, which decay in lepton pairs of opposite charge, have been extensively investigated using this analysis technique.
Dilepton spectra can be even employed to measure $\text{c}\overline{\text{c}}$ and $\text{b}\overline{\text{b}}$ production cross sections in pp, p--A and A--A collisions. In pp collisions, the measurement of $\text{c}\overline{\text{c}}$ and $\text{b}\overline{\text{b}}$  cross sections through the study of dielectron production was reported at the Large Hadron Collider (LHC) by the ALICE collaboration at $\sqrt{s}$ = 13 TeV \cite{ALICE:2018gev}, while at the Relativistic Heavy Ion Collider (RHIC), the PHENIX collaboration carried out similar measurements in pp collisions at $\sqrt{s}$ = 200 GeV, exploring both dielectron \cite{PHENIX:2008qav} and dimuon \cite{PHENIX:2018dwt} production.

Thanks to the capabilities of the ALICE detector, it is possible to reconstruct dileptons not only in the electron channel at midrapidity (|\textit{y}| < 0.9) but also in the muonic channel at forward rapidity (2.5 < \textit{y} < 4) with the muon spectrometer.
The measurement of heavy-flavor cross sections inspecting the continuum regions of the opposite sign (OS) dimuon invariant mass spectrum in pp collisions at $\sqrt{s}$ = 13 TeV represents an unexplored complementary measurement with respect to the dielectron studies, since it allows to cover a different rapidity range. Furthermore, these measurements will allow investigating the charm and beauty quark production at higher energies with respect to those accessible at RHIC.      
The ideal invariant mass regions to perform this measurement are $\text{4} < m_{\mu^{+}\mu^{-}} <\text{ 9 GeV}/c^2$ and $m_{\mu^{+}\mu^{-}}>$ 11 $ \text{GeV}/c^2$ since they are mostly populated by the semileptonic decays of open heavy-flavor hadrons, i.e. bound states of a heavy quark combined with up, down, or strange quarks, produced by the hadronization of $\text{c}\overline{\text{c}}$ and $\text{b}\overline{\text{b}}$ pairs. 
An additional contribution, mainly in the low dimuon mass regions, comes from the decay of light-flavor (LF) particles, especially $\pi$ and K mesons, which contribute to the combinatorial background.

The goal of the study presented in these proceedings is a first comparison between the dimuon invariant mass spectrum measured with the ALICE detector in pp collisions at $\sqrt{s} = \text{13 TeV}$, based on an integrated luminosity $\mathcal{L}_{\text{int}} \sim 25$ pb$^{-1}$, and predictions of the different contributions to the dimuon spectrum obtained with the PYTHIA8 Monte Carlo (MC) generator \cite{Sjostrand:2014zea}. 
    
\section{Analysis}

The characteristics and the performances of the ALICE detector are presented in detail in Ref. \cite{ALICE:2014sbx}. The study presented in this contribution is carried out in the forward rapidity region, 2.5 $ < y < $ 4, corresponding to the coverage of the ALICE muon spectrometer. During the LHC Run 2 data taking, the spectrometer was composed of: a) a system of absorbers, necessary to filter the background from hadrons ($\pi$, K) ; b) a dipole magnet, which bends muon trajectories to measure their momenta; c) the muon tracker, which consists of a set of tracking chambers placed along the beam line inside, as well as in front of and behind the dipole magnet; d) the muon trigger, composed by four trigger chambers, placed behind a passive iron wall.

The analysis technique consists in extracting the probability density functions (PDFs) of the various components to the dimuon yield in the continuum regions from the transverse momentum ($p_{\text{T}}$) and invariant mass ($m$) distributions simulated with PYTHIA8. The corresponding distributions are then fitted to data with a template obtained from the superposition of these PDFs properly weighted.

The heavy and light flavor contributions dominating the dimuon yield in the invariant mass regions of interest were studied with two dedicated MC simulations, both using PYTHIA8 with the Monash tune \cite{Skands_2014}. With a minimum bias (MB) simulation, the magnitude of the light-flavor background and heavy-flavor component in the two continuum regions were studied. In order to investigate the different components of the dimuon invariant mass spectrum originating from the decay of HF particles, a heavy-flavor enriched simulation was employed. The latter MC was obtained by selecting events which contain at least one heavy quark pair generated and the presence of at least one muon track from a heavy-flavor particle decay reconstructed between 2.5 < \textit{y} < 4. 
Then, the charm and beauty cross sections can be measured by fitting the distributions to data, according to the following calculation:
\begin{equation}
\sigma^{\text{meas}}_{\text{c}\overline{\text{c}},\text{b}\overline{\text{b}}}=\frac{N^{{\text{c}\overline{\text{c}},\text{b}\overline{\text{b}}}}_{\mu \mu, \text{\text{fit}}} }{N^{{\text{c}\overline{\text{c}},\text{b}\overline{\text{b}}}}_{\mu \mu, \text{PYTHIA}}}\times \sigma^{\text{PYTHIA}}_{\text{c}\overline{\text{c}},\text{b}\overline{\text{b}}},
\end{equation}
where $N^{{\text{c}\overline{\text{c}},\text{b}\overline{\text{b}}}}_{\mu \mu, \text{fit}} $ and $N^{{\text{c}\overline{\text{c}},\text{b}\overline{\text{b}}}}_{\mu \mu, \text{PYTHIA}} $ are the numbers of dimuons produced by the decays of charm and beauty particles extracted with a template fit over the data, and predicted by PYTHIA8 in the HF enriched simulation, respectively. The term $\sigma^{\text{PYTHIA}}_{\text{c}\overline{\text{c}},\text{b}\overline{\text{b}}}$ is the charm and beauty cross sections in the rapidity region  2.5 < \textit{y} < 4, obtained as:
\begin{equation}
\sigma^{\text{PYTHIA}}_{\text{c}\overline{\text{c}},\text{b}\overline{\text{b}}}= \frac{N^{2.5 < y < 4}_{\text{c}\overline{\text{c}},\text{b}\overline{\text{b}}, \text{MC}}}{N_{\text{Ev}, \text{MC}}}\times \sigma_{\text{PYTHIA}}
\end{equation}
where $N^{2.5 < y < 4}_{\text{c}\overline{\text{c}},\text{b}\overline{\text{b}}, \text{MC}}$ represents the number of $\text{c}\overline{\text{c}}$ and $\text{b}\overline{\text{b}}$ quark pairs produced in the rapidity coverage of the ALICE muon spectrometer according to PYTHIA8; $N_{\text{Ev}, \text{MC}}$ is the number of Monte Carlo events generated; $\sigma_{\text{PYTHIA}}$ is the inelastic pp cross section employed by PYTHIA8 in the simulation.
The quantities $N^{{\text{c}\overline{\text{c}},\text{b}\overline{\text{b}}}}_{\mu \mu, \text{fit}} $ are evaluated with a simultaneous fit to the reconstructed dimuon $p_{\text{T}}$ and \textit{m} distributions measured by ALICE, with a template obtained from the sum of three contributions: the PDF for dimuons coming from the decays of two charm particles, the PDF for dimuons originating from two beauty particles (including the case in which a charm hadron is produced via a b-hadron decay) and the PDF corresponding to the mixed case (i.e. with the dimuon composed by one muon originating from a charm hadron decay and the other coming from a beauty hadron decay). 
The PDFs were extracted by fitting the reconstructed spectra from the HF enriched simulation, normalized to their integrals, by using phenomenological functions.

\subsection{Contribution to the dimuon yield}

\begin{figure}[htb]
\centering
\begin{subfigure}{.495\textwidth}
    \centering
    \includegraphics[width=\linewidth,height=6.25cm]{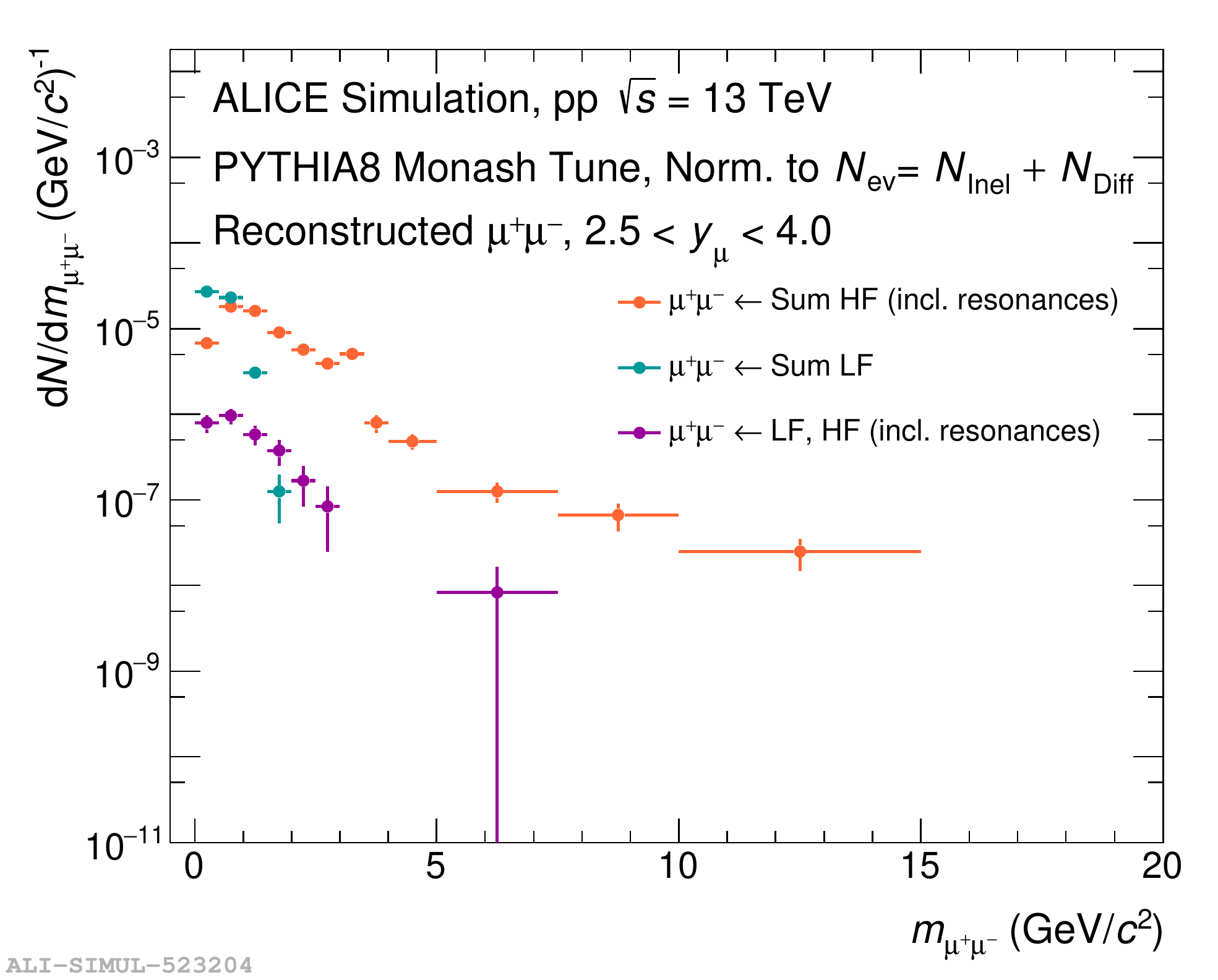}
    \caption{}
    \label{subfg1.1}
\end{subfigure}
\begin{subfigure}{.495\textwidth}
  \centering
  \includegraphics[width=\linewidth,height=6.25cm]{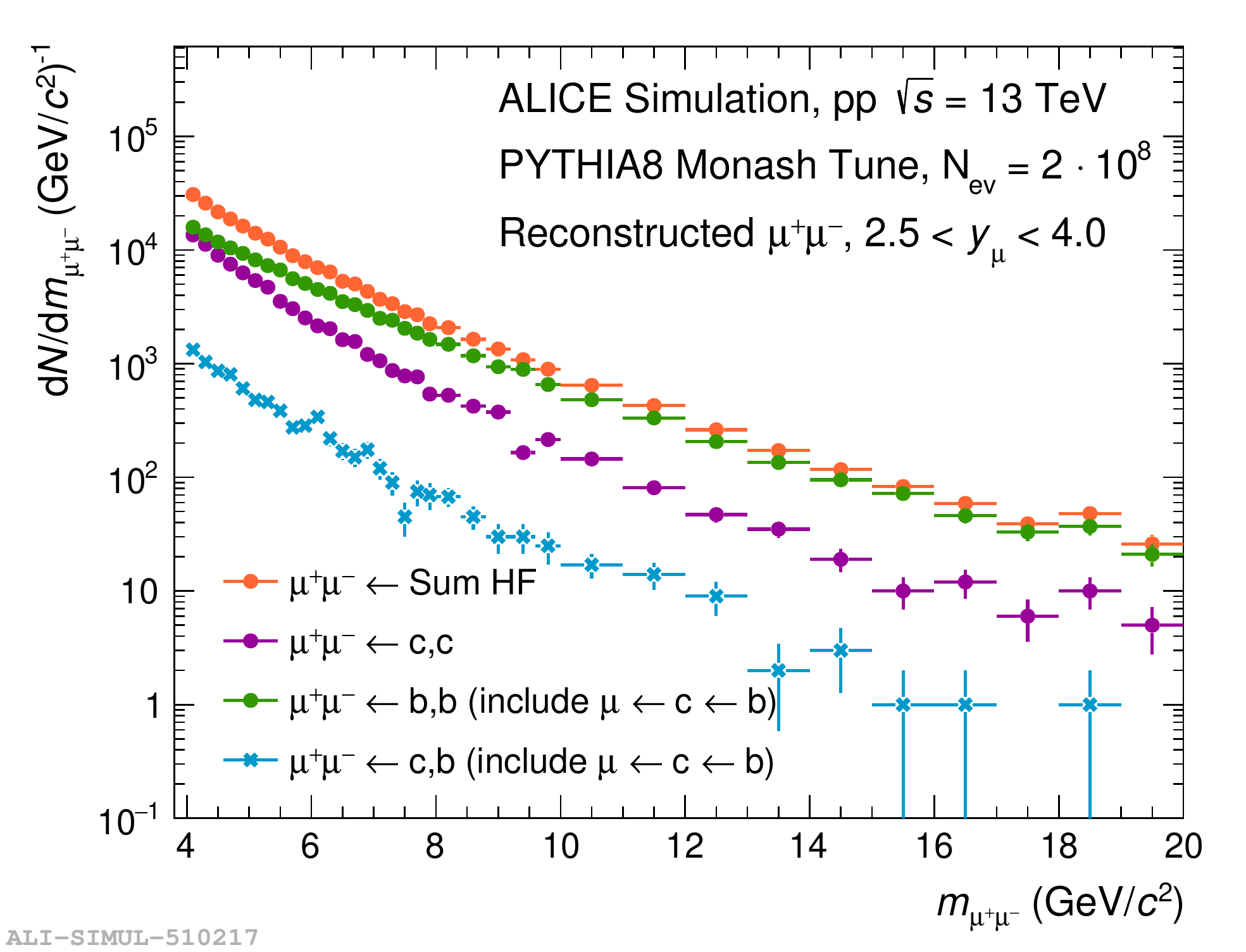}
  \caption{}
  \label{subfg1.2}
\end{subfigure}
\caption{OS dimuon invariant mass distributions from the Monte Carlo simulations \cite{Sjostrand:2014zea}. In the left panel (\textbf{a}) the LF, HF and mixed LF,HF distributions are shown, normalized to the number of events simulated. The right panel (\textbf{b}) displays the three components (charm, beauty and mixed charm and beauty) to the total number of dimuons from HF (Sum HF), as obtained from the HF enriched Monte Carlo, when the cut $m_{\mu^{+}\mu^{-}}>\text{4}$ $\text{GeV}/c^{2}$ is applied.}
\label{fg1}
 \end{figure}

In Fig. \ref{fg1} (a), the dimuon invariant mass spectra obtained from the MB simulation are reported. The light-flavor (Sum LF) contribution is negligible for $m_{\mu^{+}\mu^{-}}>\text{3}$ $\text{GeV}/c^{2}$. In the same mass region, the contribution of combined muons from light-flavor and heavy-flavor decays (LF, HF), is one order of magnitude smaller with respect to the heavy-flavor contribution (Sum HF). 

\subsection{Template fit to the data}
\begin{figure}[htb]
\centering
\begin{subfigure}{.495\textwidth}
    \centering
    \includegraphics[width=\linewidth,height=6.25cm]{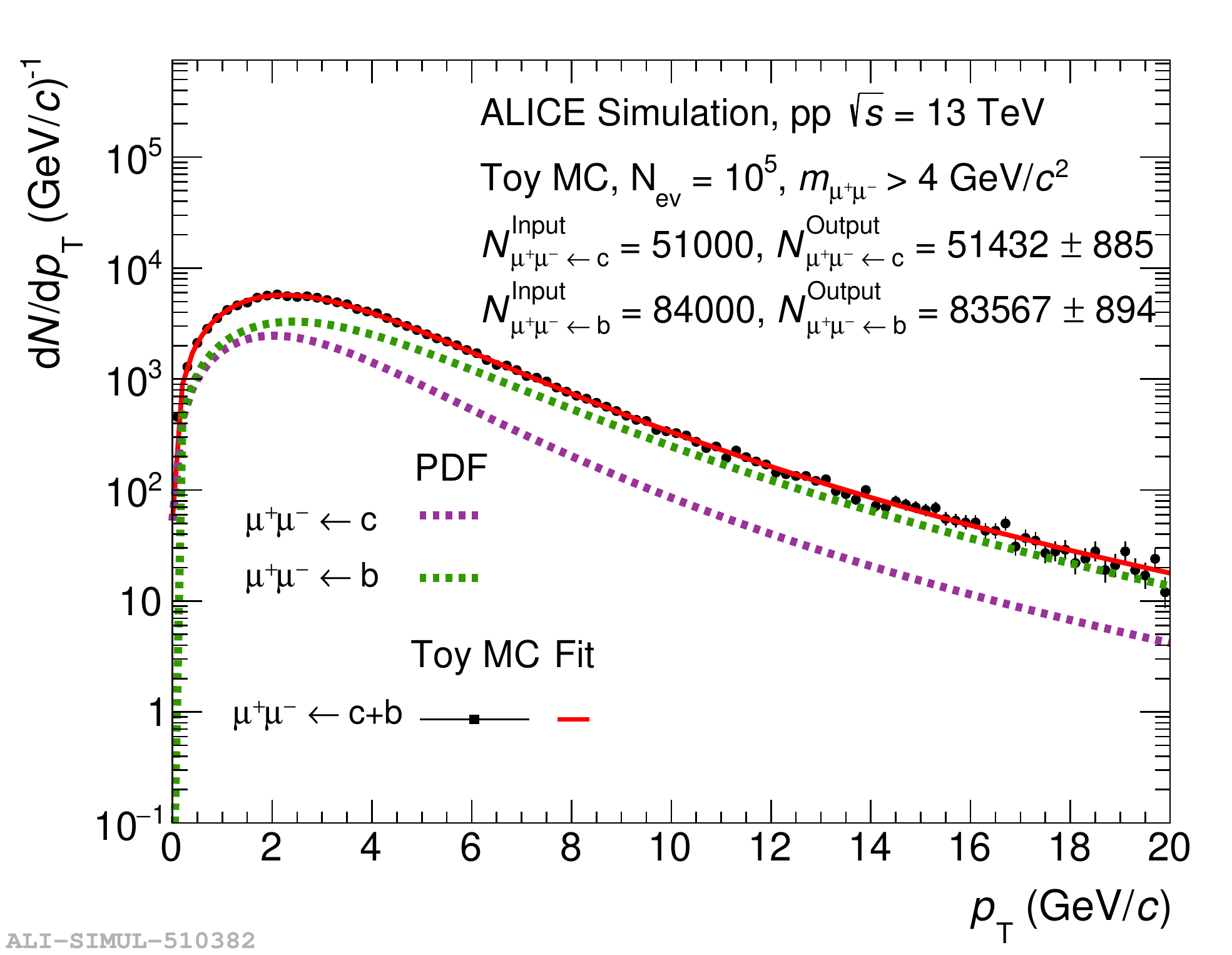}
    \caption{}
    \label{subfg2.1}
\end{subfigure}
\begin{subfigure}{.495\textwidth}
  \centering
  \includegraphics[width=\linewidth,height=6.25cm]{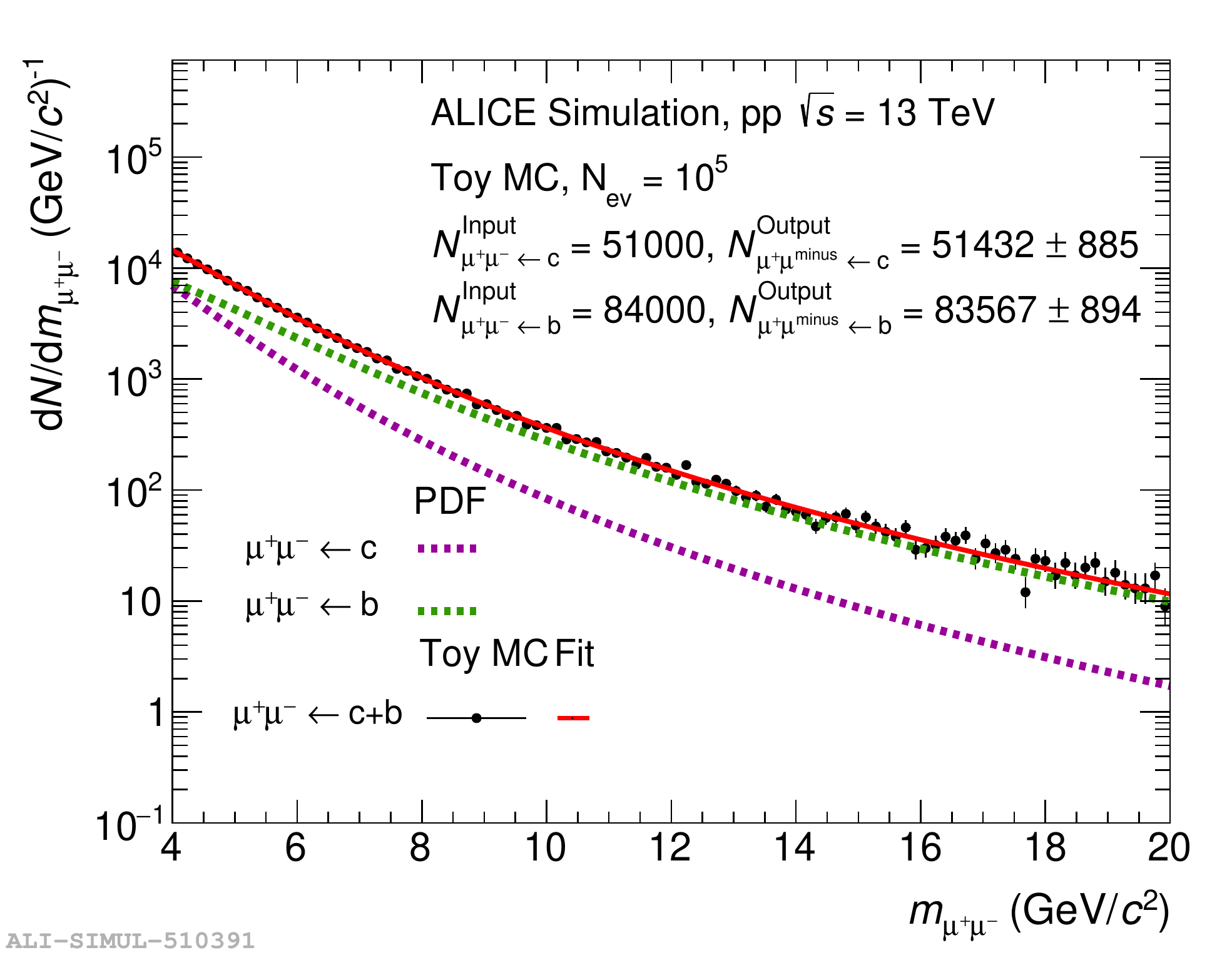}
  \caption{}
  \label{subfg2.2}
\end{subfigure}
\caption{Results of the closure test made to verify the goodness of the PDF extraction for the charm and beauty components of the HF dimuon yield. The toy Monte Carlo was fitted in $p_{\text{T}}$ (\textbf{a}) and \textit{m} (\textbf{b}) simultaneously.}
\label{fg2}
 \end{figure}

According to PYTHIA8 calculations, the selection $m_{\mu^{+}\mu^{-}}>\text{4}$ $\text{GeV}/c^{2}$, although removing 97\% of the total reconstructed dimuon yield, is expected to be extremely effective in getting a sample free of the light-flavor background, in which heavy-flavor particle decays represent the main dimuon source. Investigating the various heavy-flavor contributions to this mass region, it can be seen in Fig. \ref{fg2} (b) that the OS dimuon yield for $m_{\mu^{+}\mu^{-}}$ > 4~ $\text{GeV}/c^{2}$ is mostly populated by the decays of beauty particles. They account for 60.5\% of the total dimuon yield, while the charm particles and the mixed contribution fractions are 35.9\% and 3.6\%, respectively. 
The PDFs of the dimuon $p_{\text{T}}$ and \textit{m} distributions produced by charm and beauty particle decays were extracted by fitting the spectra from the HF enriched MC simulation with phenomelogical fit functions. To test the validity of this procedure a closure test were conducted by creating, via a toy MC, $p_{\text{T}}$ and \textit{m} dimuon distributions by combining the shapes extracted from the HF enriched MC, each one weighted with $N^{{\text{c}\overline{\text{c}},\text{b}\overline{\text{b}}}}_{\mu \mu, \text{PYTHIA}}$. Hence, the toy pseudo-data were fitted simultaneously in $p_{\text{T}}$ and \textit{m} by the sum of the two PDFs keeping the normalizations of the charm and beauty components as free parameters. In Fig. \ref{fg2} the outcome of the fit test is reported, in particular projections of the fit results over $p_{\text{T}}$ and invariant mass distributions are shown in Fig. \ref{fg2} (a) and \ref{fg2} (b), respectively. The results confirm the absence of a bias in the PDF extraction given the compatibility within uncertainties, between the $N^{{\text{c}\overline{\text{c}},\text{b}\overline{\text{b}}}}_{\mu \mu, \text{PYTHIA}}$, used as input to create the two toy distributions, and the parameters $N^{{\text{c}\overline{\text{c}},\text{b}\overline{\text{b}}}}_{\mu \mu, \text{fit}} $ retrieved from the fit. This test is crucial to confirm the validity of the fitting procedure foreseen for the real data.

\section{Conclusions}

In these proceedings, the analysis technique to measure the $\text{c}\overline{\text{c}}$ and $\text{b} \overline{\text{b}}$ cross sections in pp collisions, exploring continuum regions of the dimuon invariant mass spectrum at forward rapidity, is presented. It represents a promising new approach for this measurement. Furthermore, the procedure to extract the PDFs for the $p_{\text{T}}$ and $m$ distributions of the dimuons from charm and beauty particle decays, as well as the closure test to verify the goodness of the extraction are reported. The future steps to achieve the measurement of charm and beauty production cross sections are the evaluation of the number of dimuons from charm and beauty decays by a template fit of the corresponding distributions in data, based on the full Run 2 statistics collected in pp collisions at $\sqrt{s}$ = 13 TeV. Furthermore, the usage of fitting templates based on different MCs, such as POWHEG \cite{Oleari_2010}, will be considered in order to perform a comparison with PTYHIA8 results.


\begin{thebibliography}{99}

\bibitem{ALICE:2018gev}Acharya, S. \textit{et al.} (ALICE Collaboration), {\em Phys. Lett. B}. \textbf{788} pp. 505-518 (2019).

\bibitem{PHENIX:2008qav}Adare, A. \textit{et al.} (PHENIX Collaboration), {\em Phys. Lett. B} \textbf{670} pp. 313-320 (2009).

\bibitem{PHENIX:2018dwt}Aidala, C. \textit{et al.} (PHENIX Collaboration), {\em Phys. Rev. D}. \textbf{99}, 072003 (2019).

\bibitem{Sjostrand:2014zea}Sjöstrand, T. \textit{et al.}, {\em Comput. Phys. Commun.} \textbf{191} pp. 159-177 (2015).

\bibitem{ALICE:2014sbx}Abelev, B. \textit{et al.} (ALICE Collaboration), {\em Int. J. Mod. Phys. A}. \textbf{29} pp. 1430044 (2014).

\bibitem{Skands_2014} Skands, P. \textit{et al.}, {\em The European Physical Journal C}. \textbf{74} (2014,8).

\bibitem{Oleari_2010} Oleari, C. , {\em Nuclear Physics B - Proceedings Supplements} \textbf{205-206} pp. 36-41 (2010,8)

\end{thebibliography}
\end{document}